%% file: IMPOTENT.tex
\documentclass[a4paper]{article}
\usepackage{graphicx}
\usepackage{graphics}
\usepackage{subfigure}
\usepackage{a4wide}

\newcommand{\isotope} [2] {\ensuremath{{}^{#2}\!{\mathrm{#1}}}}
\newcommand{\Cone}{\ensuremath{\mathrm{C1}}}
\newcommand{\Ctwo}{\ensuremath{\mathrm{C2}}}
\newcommand{\Tone}{\ensuremath{\mathrm{T_1}}}
\newcommand{\h}{\ensuremath{\mathrm{H}}}
\def        \eps           {\varepsilon}
\def        \sec           {\ensuremath{\mathrm{s}}}
\def        \UdeM    {1}
\def        \TecChem {2}
\def        \Polytec {3}
\def        \Oxford  {4}
\def        \TecCS   {5}
\def        \TecPhys {6}
\def\blfootnote{\xdef\@thefnmark{}\footnotetext}

\title{Experimental Heat-Bath Cooling of Spins}

\author
{G.~Brassard$^{\UdeM}$\hskip -0.25ex
\footnote{The authors are listed according to the alphabetical order.},
Y.~Elias$^{\TecChem}$, J.~M.~Fernandez$^{\Polytec}$,
H.~Gilboa$^{\TecChem}$,\\ J.~A.~Jones$^{\Oxford}$,
T.~Mor$^{\TecCS}$, Y.~Weinstein$^{\TecPhys\dagger}$ and L.~Xiao$^{\Oxford}$\\
\\
\normalsize{$^{\UdeM}$D{\'e}partement IRO, Universit{\'e} de Montr{\'e}al, Montr{\'e}al, H3C 3J7~~Canada}\\
\normalsize{$^{\TecChem}$Department of Chemistry, Technion 32000, Haifa, Israel}\\
\normalsize{$^{\Polytec}$D{\'e}partement de g{\'e}nie informatique,}\\
\normalsize{{\'E}cole Polytechnique de Montr{\'e}al, Montr{\'e}al, H3C 3A7~~Canada}\\
\normalsize{$^{\Oxford}$Center for Quantum Computation, Clarendon Laboratory,}\\
\normalsize{University of Oxford, Parks Road, Oxford OX1 3PU, United Kingdom}\\
\normalsize{$^{\TecCS}$Department of Computer Science, Technion 32000, Haifa, Israel}\\
\normalsize{$^{\TecPhys}$Department of Physics, Technion 32000, Haifa, Israel}\\
\\
\normalsize{$^\dagger$To whom correspondence should be addressed. E-mail:
yossiv@technion.ac.il}\\
}

\begin{document}

\maketitle

\blfootnote{Abbreviations: AC, algorithmic cooling; NMR, nuclear magnetic
resonance; SNR, signal to noise ratio; PT, polarization transfer; POTENT, polarization
transfer via environment thermalization}

% Start the footnote count from three as we used the one for stating the 
% alphabetical order and another one for contact details for the 
% corresponding author.
\setcounter{footnote}{2}
\begin{abstract}
%
% FOR PNAS (250 words total are allowed)
%
Algorithmic cooling is a novel technique to generate ensembles of
highly polarized spins, which could significantly improve the
signal strength in Nuclear Magnetic Resonance (NMR) spectroscopy.
It combines reversible (entropy-preserving) manipulations and
irreversible controlled interactions with the environment, using
simple quantum computing techniques to increase spin polarization
far beyond the Shannon entropy-conservation bound. 
Notably, thermalization is beneficially employed as an integral 
part of the cooling scheme, contrary to its ordinary destructive 
implications.
 We report the first cooling experiments bypassing Shannon's 
entropy-conservation  
bound, performed on a standard liquid-state NMR spectrometer. We believe 
that this experimental success could pave the way for the first near-future
application of quantum computing devices.
%Total 111
\end{abstract}

%
% Parag: MRS
%
\section*{Introduction}
An efficient technique to generate ensembles of highly-polarized
spins is a Holy Grail for Nuclear Magnetic Resonance (NMR)
spectroscopy
and imaging (MRI).  
The resulting enhancement of the signal-to-noise ratio (SNR)
permits faster data acquisition, or reduces the amount of material
required, thus enabling more efficient analysis of chemicals and
visualization of tissues.
Methods to
overcome the low sensitivity of nuclear magnetic resonance
~\cite{Ernst,Slichter} include high magnetic fields (limited to about
20\,T), signal averaging (time consuming), and
temperature reduction (impractical for many samples). Another
solution is to cool the spins without cooling the environment:
spin-half particles in magnetic fields have steady-state
polarization biases inversely proportional to the temperature and
so spins exhibiting polarization biases in excess of the
equilibrium bias are considered \emph{cool}, even when their
environment is warm
~\cite{Slichter}. Such spin-cooling is an important method of
increasing SNR: at room temperature in a constant magnetic field,
a polarization enhancement by a factor of $M$ improves the SNR by the
same factor, and reduces the required signal averaging time by a factor of 
$M^2$.

%
% Parag: spin cooling refs.
%
Spin-cooling techniques are not a new concept. Many NMR studies
rely on a variety of methods to transfer polarization from high bias spins to
low bias spins (see~\cite{Slichter,Sorensen89} and references therein) increasing the
latter's signal intensity. Alternatively, several more recent approaches are
based on the creation of very high polarizations, e.g.~dynamic nuclear
polarization~\cite{Larsen03}, \emph{para}-hydrogen in two-spin
systems~\cite{Anwar04}, and hyperpolarized xenon~\cite{Oros04}. In
addition, there are other spin-cooling methods that are based on
general unitary transformations~\cite{Sorensen89}, and on (closely
related) data compression methods in closed systems~\cite{SV99}. A
different effective-cooling method,
\emph{algorithmic cooling} (AC), makes use of entropy manipulations in an
open system~\cite{BMRVV,FLMR04}. These entropy manipulation
techniques in closed and open systems employ simple NMR quantum
computing tools.

\begin{figure}[tb]
\begin{center}
\includegraphics[scale=1]{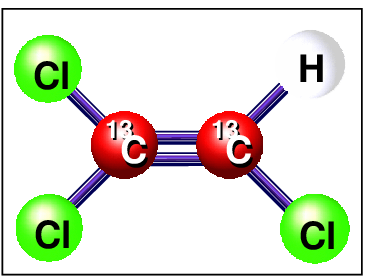}
\end{center}
\caption{
Trichloroethylene labeled with two \isotope{C}{13}. We
denote the leftmost \isotope{C}{13} in this figure as \Cone\ and
the other, neighboring \isotope{H}{1}, as \Ctwo . In our
experiments, the resonance frequencies were 125.773354, 125.772450
and 500.133245~MHz for \Cone , \Ctwo\ and \h\ respectively.
The scalar coupling constants were 201, 103 and 9 Hz
between \Ctwo-\h, \Cone-\Ctwo, and \Cone-\h, respectively, while
\Tone\ relaxation times were measured at $43\pm 4.0\sec$ and
$20\pm 2.0\sec$ for \Cone\ and \Ctwo\ respectively, and $3.50\pm
0.05\sec$ for \h.
}
\label{fig:TCE}
\end{figure}

%
% Parag: NMRQC
%
Nuclear Magnetic Resonance Quantum Computing (NMRQC)~\cite{Jones, Jones00,
Yehuda, Vandersypen04} was proposed and demonstrated in
1996~\cite{Cory96,CFH97,Ger-Chuan97}. NMRQC researchers have
subsequently implemented several quantum algorithms, using
standard NMR equipment, on spin systems (molecules) containing up
to 7 quantum bits (that is, two-level quantum systems, also called
qubits)~\cite{Factoring-15}.
The system is commonly a solution containing an ensemble of
identical molecules. Each molecule acts as an independent quantum
computer, and the same algorithm is run in parallel on all
computers. A qubit is represented by a spin-half nucleus in each molecule.
For instance, trichloroethylene (TCE) molecules (see figure~\ref{fig:TCE}) with two
\isotope{C}{13} nuclei and one \isotope{H}{1} nucleus provide three-qubit
registers.
Gate operations are implemented by sequences of radio frequency
pulses, as in conventional NMR. Two key aspects of such
ensemble quantum computing are the highly mixed initial state and
the fact that when measuring a qubit one obtains a sum of values
for that qubit, averaged over the ensemble.
The application of conventional quantum algorithms to such mixed-state computers
suffers from a severe scalability
problem~\cite{Ger-Chuan97,Warren,SV99,Jones00,BMRVV}.
%
% Parag: REM - reversible entropy manipulations
%
To resolve this problem, and to prove the potential scalability
of NMRQC, Schulman and Vazirani~\cite{SV99} suggested the use
of reversible (entropy preserving) spin-cooling schemes.
The practicality of such unitary schemes~\cite{Sorensen89,SV99}
is highly limited due to the Shannon  
entropy-conservation bound 
and the S{\o}rensen unitarity 
bound. For instance, the spin temperature of a single spin (qubit)
in an $n$-qubit molecule (at room temperature) cannot be decreased
by more than a multiplicative factor of $\sqrt{n}$ due to 
the entropy-conservation bound (namely, 
Shannon's bound on reversible entropy manipulations\footnote{
The total
entropy of such a molecule,   
$H(n) \sim n(1-\epsilon^2/\ln 4)$,
is compressed so that $n-1$ spins have full entropy; the remaining
spin satisfies 
$H({\rm single})\ge 1-(\sqrt{n}\epsilon)^2/\ln4$.
} --- the source coding
theorem~\cite{Ash90,CT91}).
 The S{\o}rensen bound is tighter than the Shannon bound as Shannon does not
 prohibit the use of non-unitary transformations. 

\section*{Algorithmic Cooling of Spins}
%
% Parag: AC
%
To increase spin polarization far beyond these two bounds, Boykin,
Mor, Roychowdhury, Vatan and Vrijen~\cite{BMRVV} suggested a novel
technique, \emph{algorithmic cooling of spins}, which combines
reversible manipulations with controlled irreversible interactions
with the environment. The reversible steps consist of compressing
entropy, then transferring it to ``reset'' spins, that is spins that
reach thermal equilibrium rapidly (qubits with very short relaxation
times). Consequently, these reset spins (reset bits\footnote{We use
the term ``bit'' and not ``qubit'', as our cooling algorithms are
classical. We refer to ``qubits'' when appropriate. Furthermore, we
use the term ``spin'' to mean a spin-half nucleus.}) thermalize and
lose their excess entropy irreversibly to the environment.

%
% Parag: practicable AC
%
More recently, Fernandez, Lloyd, Mor, and Roychowdhury~\cite{FLMR04}
have designed improved cooling algorithms that can be applied to
\emph{short molecules}, thus greatly expanding the potential for
near-future application to high-sensitivity magnetic resonance
spectroscopy; a polarization improvement by a factor of about
$(3/2)^{(n-1)/2}$ on a single spin is obtained using a simple
algorithm applied
to $n$ spins (of which one is a reset spin). For more details, and
several interesting algorithms and bounds see~\cite{FLMR04, SMW05}.

%
% Parag: AC experimentation
%
The reversible manipulations required have already been
demonstrated experimentally~\cite{Sorensen89,BCS-exp}, while the
irreversible step of AC has not been previously reported. Here we
show experimentally, for the first time, controlled entropy
extraction from a spin system, to bypass Shannon's entropy-conservation
bound.
To do so, we implement a restricted form of AC called \emph{heat-bath
cooling}, where polarization transfer steps (but no compression steps)
are combined with controlled interactions with the environment (reset
steps) to achieve this entropy reduction.
This experimental implementation presents a major step in the development of
open-system spin-cooling techniques, and
therefore paves the road for the full 3-bit algorithmic cooling process,
and for more dramatic cooling as suggested in~\cite{BMRVV,FLMR04,SMW05}.

\section*{Materials and methods}
In our AC experiment we employ the 3-qubit heteronuclear molecule
TCE, shown in Figure~\ref{fig:TCE}, in which the hydrogen has
relatively fast relaxation and can function as a reset bit.
Furthermore, at thermal equilibrium, the polarization of both
\isotope{C}{13} nuclei is approximately the same, say $\eps$, and
that of the proton is very close to $4\eps$. Therefore it is also
useful for conventional (entropy preserving) polarization transfer
(PT). When PT is combined with fast relaxation (reset) of the proton
and with polarization compression, cooling of TCE beyond Shannon's
and S{\o}rensen's bounds can be achieved in various ways, based on
algorithms described in~\cite{FLMR04,SMW05}, positioning this
molecule as a good testbed for various cooling experiments. 
The
bound we bypass in this work is the Shannon bound regarding the 
conservation of the {\em full
entropy} of the 3-bit system. The Shannon entropy of the system at
thermal equilibrium is approximately $H\sim 3-(4^2 + 1^2 +
1^2)\eps^2/\ln 4 = 3 - 18\eps^2/\ln 4$, in \emph{bit} units\footnote{The
$\ln 4$ emerges from the $2^{\mathrm{nd}}$ order term in the Taylor
approximation.}.
The \emph{information content} of the molecule, that is the
difference from full entropy, is given in this case by $I=3-H\sim
18\eps^2/\ln 4$. Here we shall be interested in increasing this
value, $I$, thus decreasing the total entropy $H$, and cooling the
entire system\footnote{Note that conventional entropy preserving
algorithms (common in data compression, and in NMR polarization
manipulations) cannot decrease the total entropy of the three bits.}.
We take advantage of two properties of
the \isotope{H}{1} in relation to the \isotope{C}{13}: greater polarization
at thermal equilibrium, and sufficiently smaller \Tone\ relaxation time.
The cooling procedure we report here is very simple, involving no
compression step.
While a system is usually heated by ``thermalization'', this partial AC
presents a simple demonstration of ``heat-bath cooling'', one of the two
building blocks of the full algorithmic cooling.
It has four steps: (1) transfer polarization
from \h\ to the far carbon~(\Cone); (2) wait for a suitable time,
$t_1$, for \h\ to repolarize; (3) transfer polarization from \h\
to the adjacent carbon~(\Ctwo); (4) wait again for a second
duration, $t_2$, for \h\ to repolarize.
 Although polarization transfer is used widely,
sometimes also combined with fast relaxation of the more polarized spins,
 this \emph{heat-bath cooling} is the first experiment
reporting bypassing Shannon's entropy-conservation bound.

%
% Parag: From SWAP to PT to INEPT
%
From an algorithmic
point of view, a transfer of polarization can be achieved by
exchanging the states of the two spins, using a SWAP gate.
The required PT, however, is unidirectional and therefore simpler than SWAP.
Implementation of such a unidirectional PT from proton to adjacent carbon
has some freedom as we do not care about any residual polarization of the
proton (after PT), which is to be reset. It is
normally inefficient to implement directly a PT gate between
non-adjacent spins due to weak scalar couplings. Step~1 above thus
comprises two sequential steps: PT(H $\rightarrow$ \Ctwo) and PT(\Ctwo
$\rightarrow$ \Cone).
A simple way of implementing PT is to use the refocused INEPT sequence,
which is an extension of the well-known INEPT sequence~\cite{Freeman79}.
Refocused INEPT
(for an illustration of the sequence see Figure~5 of supporting information)
may be considered as a unidirectional SWAP, in which the polarization of the
proton is fully transferred to the carbon.
We refer to this practical implementation of \emph{cooling by
thermalization} as POTENT: POlarization Transfer via ENvironment Thermalization.

%
% Parag: Summary in the ideal case
First consider an ideal case, where the \Tone\ ratio between each
\isotope{C}{13} nucleus and
 \isotope{H}{1} is infinite, the resonance frequencies
are in exact ratios of 1:1:4, all SWAP gates are implemented
perfectly and the long-lived spins are not prone to relaxation or
errors. Starting from equilibrium biases $\{\eps,\eps,4\eps\}$ scaled
up by $\eps$ and denoted as $\{1,1,4\}$ for \Cone, \Ctwo\ and \h,
respectively, this would result in the following sequence of polarizations:
\[
\begin{array}{rrl}
\{1,1,4\} \stackrel{PT}{\longrightarrow}  \{1,4,1\}
&\stackrel{PT}{\longrightarrow}   \{4,1,1\}
&\\
\stackrel{reset}{\longrightarrow} \{4,1,4\}
&\stackrel{PT}{\longrightarrow}   \{4,4,1\}& \stackrel{reset}{\longrightarrow}   \{4,4,4\}
\end{array}
\]
which would result in a final information content of
$$I=(4^2+4^2+4^2)\eps^2/\ln 4=48\eps^2/\ln 4,$$ a more than two-fold
increase, which clearly bypasses the entropy-conservation bound.

%
% Paragraph: The crude reality....
%
However, one must account for the finite ratio in \Tone\ values.
We performed a numerical simulation
of the POTENT pulse sequence
 using a standard relaxation model and \Tone\ relaxation rates
measured in the laboratory, to obtain the expected final bias
values of each spin. The same simulation model also provided
optimal values of the two free parameters in the experiment, $t_1$
and $t_2$, the first and second \h\ repolarization delays,
which maximize the final information content
(see details of the simulation method in supporting information).

The POTENT experiments were initially performed at
Universit{\'e} de Montr{\'e}al and later on at the Technion and Oxford.
 In some of the experiments
a paramagnetic relaxation agent was added to the sample to increase
the ratios of \Tone\ relaxation times between the \isotope{H}{1}
and both \isotope{C}{13}, as suggested in~\cite{FMW03, FMW05}.
See supporting information for details regarding the relaxation agent.
The simulation and experimental results presented in this paper
correspond to a sample that contains this relaxation agent.

\section*{Results}
\begin{figure}[h]
\begin{center}
\subfigure[][]{%
  \includegraphics[scale=0.25,angle=270]{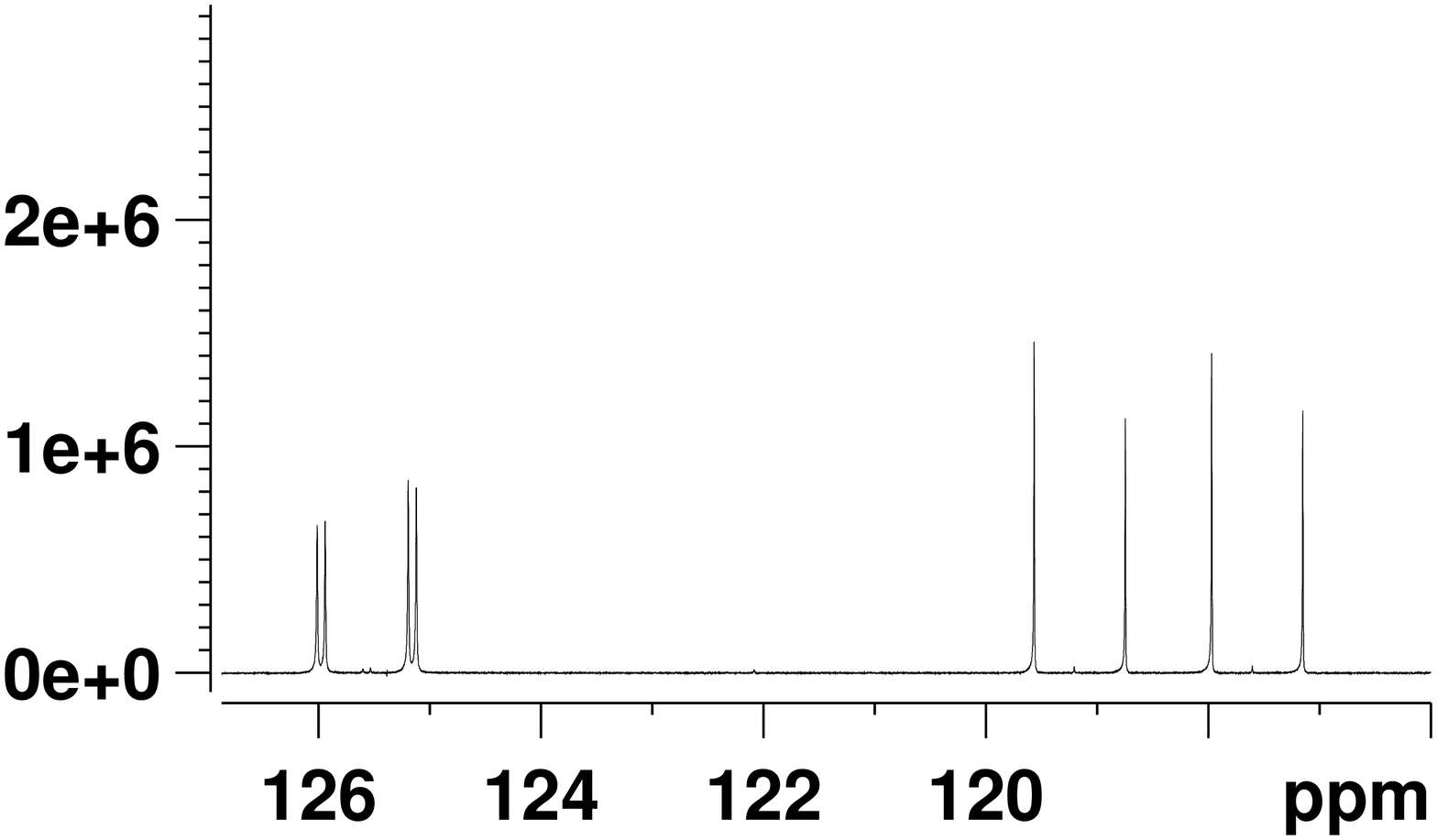}
  \label{fig:zg-C}}
\subfigure[][]{%
  \includegraphics[scale=0.25,angle=270]{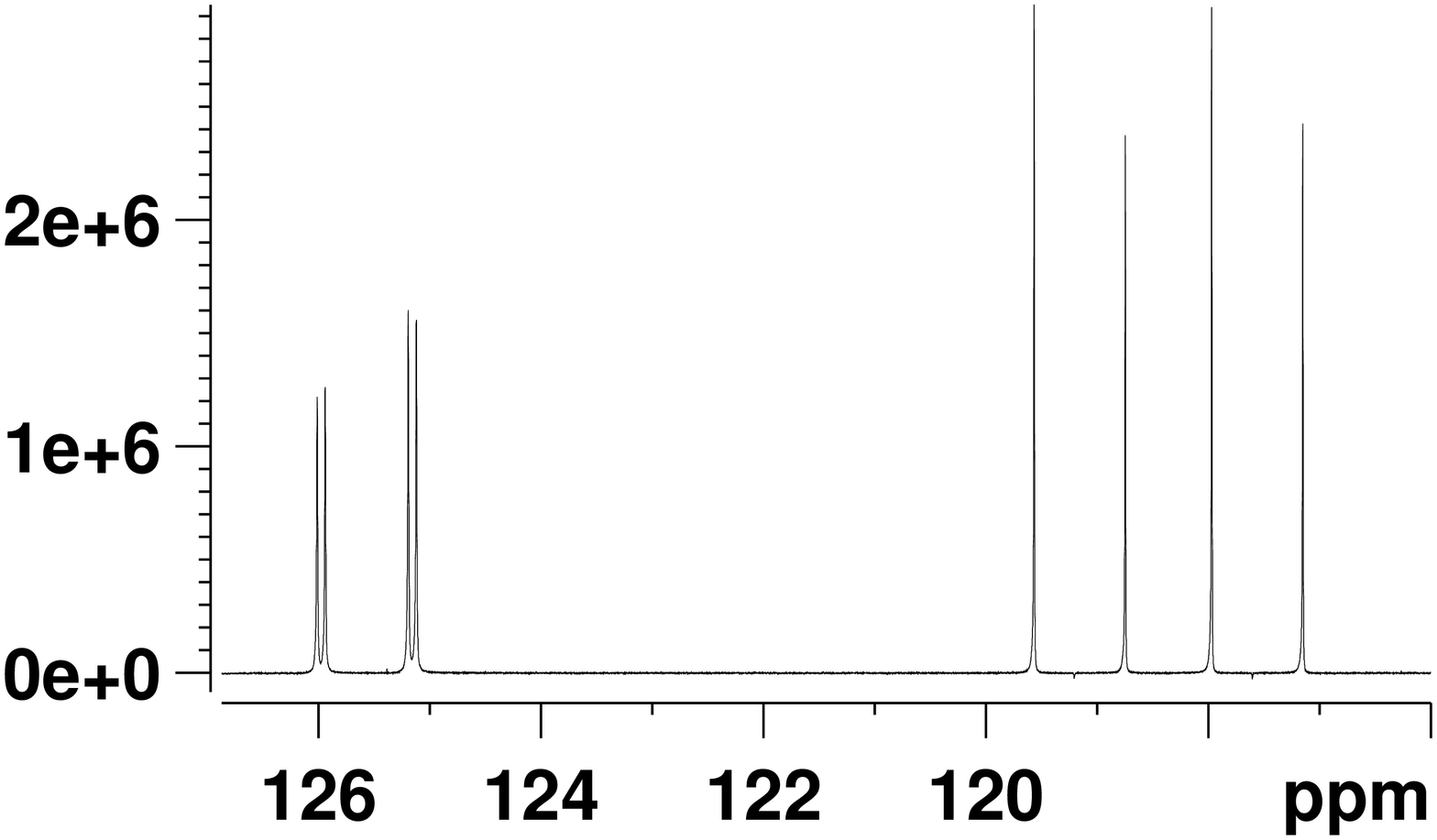}
  \label{fig:after-C}}\\
\subfigure[][]{%
  \includegraphics[scale=0.25,angle=270]{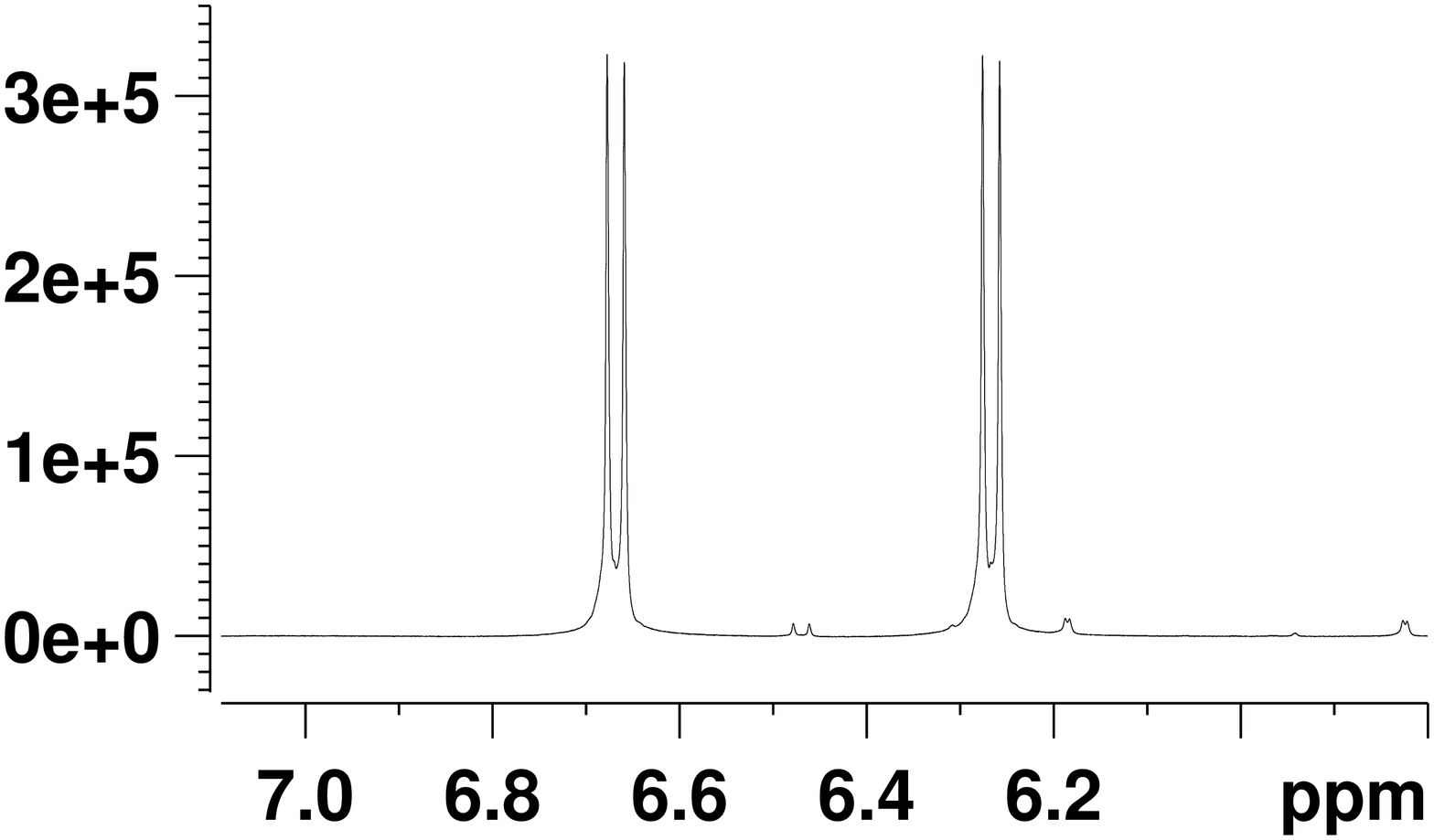}
  \label{fig:zg-H}}
\subfigure[][]{%
  \includegraphics[scale=0.25,angle=270]{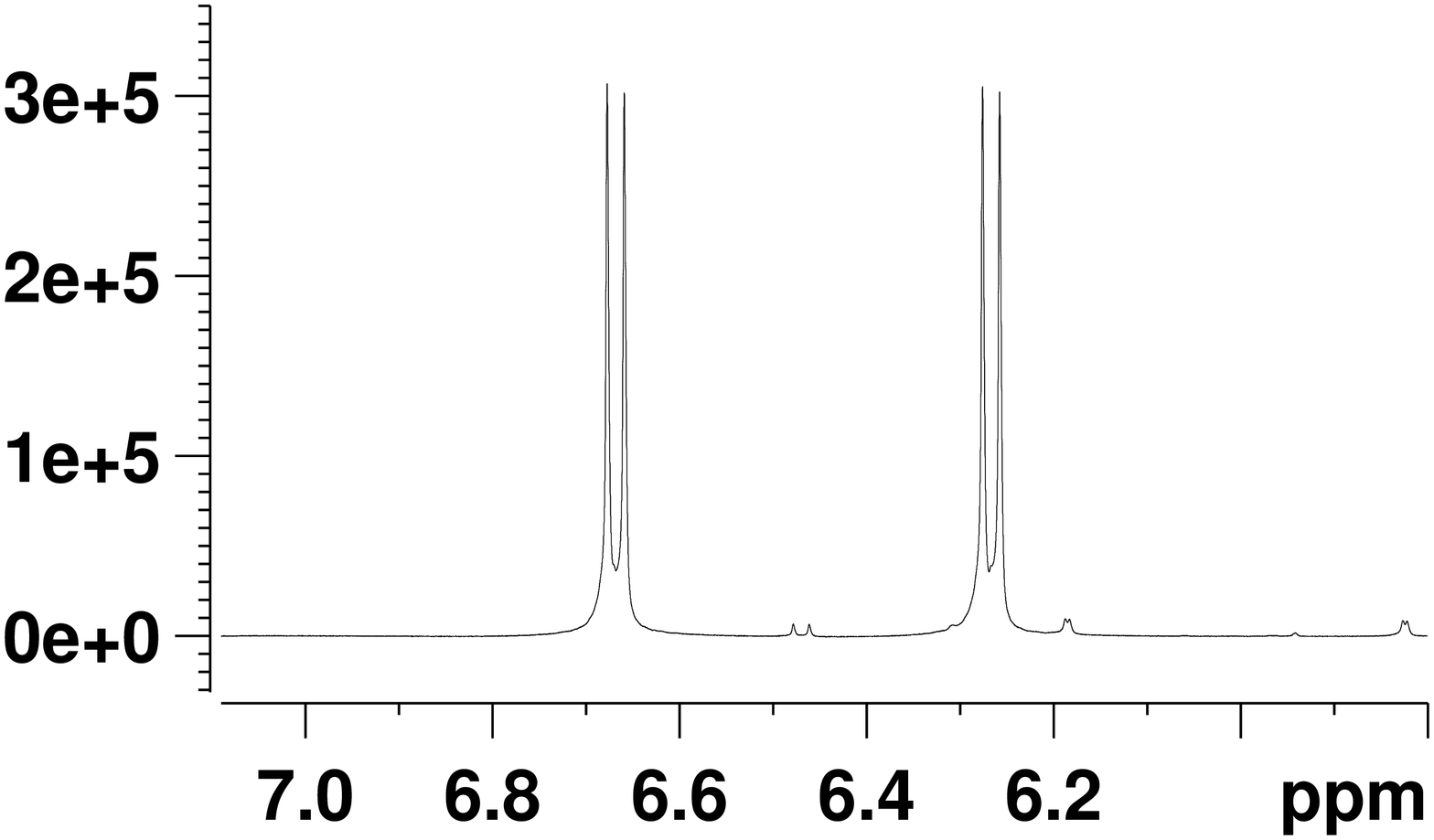}
  \label{fig:after-H}}
\caption{
  Spectra of TCE before and after the AC experiment.  Figs.\
  (a) and (b) are the \isotope{C}{13}
  spectra before and after the experiment, respectively, with the
  left multiplet being \Cone\ and the right one \Ctwo.  Figs.\
  (c) and (d) are the corresponding \isotope{H}{1} spectra
  before and after the experiment, respectively. The spectrum in
  Fig. (d) was obtained by running the AC experiment
  a second time with the exact same parameters as in
  Fig.~(b), this time observing the \isotope{H}{1} instead of
  the \isotope{C}{13}s by reversing the spectrometer channels.
}
\label{fig:before-after}
\end{center}
\end{figure}

We aqcuired spectra of TCE at equilibrium and after the cooling pulse sequence.
Figure~\ref{fig:before-after} displays \isotope{C}{13} and \isotope{H}{1}
NMR spectra for TCE. Figures \ref{fig:zg-C} and~\ref{fig:zg-H} were
obtained at thermal equilibrium and serve as a reference point for
\isotope{C}{13} and \isotope{H}{1}, respectively.
 For both \isotope{C}{13} nuclei, the increase in polarization
 can be observed by looking at the noticeably higher peaks
 compared with the reference spectrum, while the
 \isotope{H}{1} intensity is only slightly reduced.
 Experimentally acquired resonance frequencies and temperature were used to 
 calculate equilibrium biases of $1 \pm 0.003$ for 
 the carbons and $3.98 \pm 0.01$ for the proton, leading to initial information
 content of $I=17.8 \pm 0.1$ (in units of $\eps^2/\ln 4$) and ideal final 
 information content of $I=47.44$.
 Precise values for the final polarization of each nucleus were
 obtained by comparing the integrals of the peaks
(see supporting information for details regarding data analysis),
as shown in Table~\ref{tab:biases}. The table also includes
 simulated final values, accounting for actual \Tone\ relaxations.
In terms of information capacity, a final value of $20.70 \pm 0.06$ was
observed, compared with $17.8 \pm 0.1$ at thermal equilibrium
(in units of $\eps^{2}/\ln 4$), resulting in an increase of $16\% \pm 1\%$.
The ideal value of $47.44$ is devoid of relaxation constraints, while 
the simulated value of $29.59$ accounts for experimental \Tone\ values;
 for details on the simulation results see supporting
information. This discrepancy between actual and simulated values can
largely be ascribed to inevitable imperfections in polarization transfer steps,
arising from many sources\footnote{A \emph{practical} simulation, which takes
 these imperfect PTs into account, yields results much closer to
our experimental 20.70,
see supporting information for more details.}. A more subtle factor is that the
relaxation model used is naive, as the spins
do not relax completely independently but rather show significant
cross-relaxation.  Because of this effect, the final observed polarizations
were lower than expected under our single-spin \Tone\
relaxation model.

\section*{Discussion}
%
% Parag: brief summary.
%
We have successfully implemented experimental cooling of a spin
system, using an essential step of Algorithmic Cooling, ``heat-bath
cooling''. 
Our POTENT experiment combined well-controlled spin
polarization transfer on three qubits and relatively fast thermal
relaxation of the ``reset bit''. We thus increased the
polarization of \textit{two} \isotope{C}{13} nuclei using
\textit{one} proton in trichloroethylene. This is the first
reported experiment bypassing Shannon's bound on entropy
manipulation and in particular on polarization enhancement\footnote{Note
that a simpler experiment with one proton and one carbon could also bypass
Shannon's entropy-conservation bound; we chose 3-spin POTENT because it is a
necessary building block for the implementation of the full AC
technique.}.
%
%
% Parag: AC (near future)
%
This also complements the previously implemented steps of
Algorithmic Cooling, polarization
compression~\cite{Sorensen89,BCS-exp}, and polarization transfer.
Thus, our experiment shows that AC is viable in practice as a
technique for increasing spin polarization in NMR. AC is readily
usable with current off-the-shelf technology and is directly
applicable to a wide range of molecules. Full AC protocols promise
very significant SNR improvement if performed on longer molecules,
if the reset bits thermalize much faster, or if AC is combined
with other techniques for improving the SNR.
It is important to compare our experiment to some common
polarization transfer techniques. Some specialized NMR techniques
(such as continuous CP, DNP, NOE or ENDOR
~\cite{Ernst,Slichter}) may be able to bypass
Shannon's entropy-conservation bound under certain circumstances. However, to
the best of our knowledge, such bypassing has not been previously
claimed in the literature
\footnote{In contrast, the easier task of bypassing S{\o}rensen's unitarity
  bound was discussed theoretically long before AC was
suggested, see~\cite{Pines}.}.
The high level of controlled quantum operations typical of NMR
quantum computing, and the potential to reach highly polarized
states without access to an initial low temperature heat-bath,
further distinguishes AC from the above NMR methods.

While AC is a classical algorithm, its implementation (via qubits)
makes use of novel tools recently developed in the evolving field
of NMR quantum computing. This algorithm holds the potential to
significantly improve NMR spectroscopy. Therefore, our
experimental success may pave the road to the first practical
application of quantum computing.

\section*{Acknowledgements}
We thank Raymond Laflamme (RL) 
for helpful discussions and especially for
participating in the initial stages of designing the experiment; we
also thank
 Yael Balasz, Sylvie Bilodeau, Jean-Christian Boileau, Nicolas Boulant,
 Camille Negrevergne and
Tan Pham Viet for useful discussions, suggestions and help in setting
up the experiments. AC is covered by US patent No.~6,873,154.
The work of GB and JMF is supported in parts by the
Natural Sciences and Engineering Research Council of Canada.
 The work of GB is also supported in parts
 by the Canada Research Chair programme and
the Canadian Institute for Advanced Research.
The work of TM, YW and YE was supported in parts by the Israeli Ministry of
Defense. The work of YE, HG, TM, and YW was supported in parts by
the Institute for Future Defense Research at the Technion. JAJ and
LX are supported by the UK EPSRC and BBSRC.
Our first successful experiment 
(see details in supporting information)
was performed at the University of Montreal
in 2002, by GB, JMF, RL, TM and YW.

\newpage

\newpage

\newpage
\begin{table}[here]
\caption{%
  Initial and final polarizations and spin temperatures for each 
 qubit in TCE for the AC experiment shown in figure~\ref{fig:before-after}.
}~\\
{\centering
 \begin{tabular}{l|l|c|l|c}
&
\parbox{6ex}{Initial bias~($\eps$)} &
\parbox{10ex}{Simulated\ bias~($\eps$)}&
\parbox{4ex}{Observed bias~($\eps$)}
& \parbox{14ex}{Spin\\ temperature~(K)}
\\\hline\hline
\Cone &
$1\pm 0.003$  &  $2.965$ & $1.74 \pm 0.01$ & $170\pm 1$
\\
\Ctwo &
$1\pm 0.003$ & $2.602$ & $1.86 \pm 0.01$ & $159\pm 1$
\\
\h &
$3.98\pm 0.01$  & $3.734$ & $3.77 \pm 0.01$ & $312\pm 1$
\end{tabular}\par}\ \\
The resonance frequency of each nucleus is used to compute
its natural bias at thermal equilibrium at the room temperature of
$295.7\pm 1.0$ K at which the experiment was run.
A maximum bias of 3.98 could be achieved for all three spins if the
\Tone\ gaps were infinite and the polarization transfers were
implemented by perfect unitary transformations.  The simulated
values were computed by taking into consideration \Tone\ relaxation
measured in the laboratory but ignoring imperfections in the
transfers.
\label{tab:biases}
\end{table}

\input{SupportingInfo.tex}
\end{document}

%% file: SupportingInfo.tex
\newpage
\appendix

\section{Supporting Information}

\begin{figure}[h]
{\centering \resizebox*{0.5\textwidth}{!}{\includegraphics[angle=270]{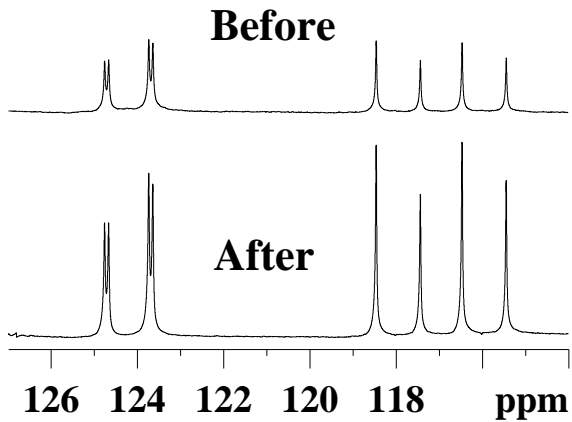}} \par}
\caption{Carbon spectra before and after 
POTENT sequence from an experiment performed at Montr{\'e}al in 2002.} 
  \label{fig:montreal}
\end{figure}

% Chronology, instruments and materials
The heat-bath cooling experiment (POTENT) 
was first performed at the Universit{\'e} de
Montr{\'e}al in March~2002 on a Bruker DMX-400.
Initial and final carbon 
spectra of the original experiment are presented 
in Figure~\ref{fig:montreal}.
Proton polarizations were calculated via the simulation model, 
as only carbon spectra were recorded. The resulting information content
showed that Shannon's entropy bound was bypassed, but we did not
have experimental results for the proton to support that conclusion.

The experiment has since been repeated at the Technion on a
Bruker Avance-500 and at Oxford University on a Varian Inova-600.
\isotope{C}{13}$_{2}$-trichloroethylene (TCE) was obtained from CDN Isotopes
(99.2\%\isotope{C}{13}) or from Cambridge Isotope Laboratories (99\%\isotope{C}{13}, diisopropylamine
stabilized).
In Montr{\'e}al and in Haifa, the same Bruker pulse sequence programs
were used on samples of TCE dissolved in deuterated chloroform
(Aldrich, 99.9\%D). In Oxford, functionally equivalent
pulse programs were used on TCE in deuterated chloroform as well as in
deuterated acetone. 
Some experiments, including the first one and those shown in this paper,
were carried out with addition of a paramagnetic relaxation reagent, 
Cr(III)acetlyacetonate, obtained from Alfa Aesar (97.5+\% pure), at a final
concentration of about 0.2 mg/mL.

\begin{figure}[p]
{\centering \resizebox*{0.88\textwidth}{!}{\includegraphics{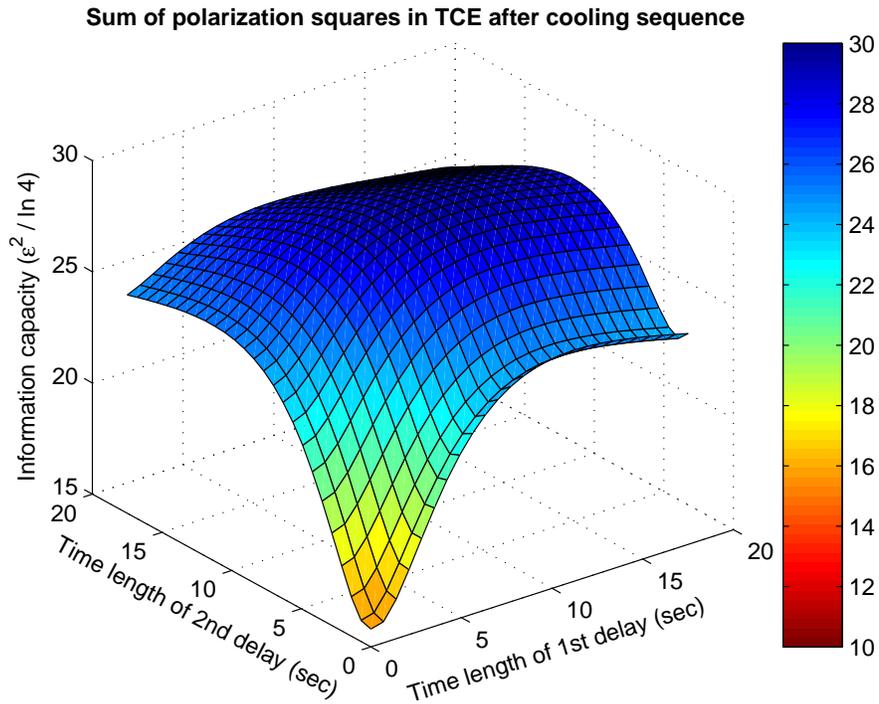}} \par}
\caption{A simulation of information capacity (IC) as a function of
  the \h\ repolarization delay times $t_1$ and $t_2$, with the 
  IC represented on the \emph{z}-axis.
  In this simulation we assume perfect polarization transfers.
  The 
  maximum IC value was numerically found at $t_1=8.25$~s and
  $t_2=9.6$~s.}
  \label{fig:matlab}
\end{figure}

% Simulation of thermalization and delay times optimization
To evaluate the expected efficiency of our AC procedure, we
wrote a Matlab program that simulates the conditions of the
TCE POTENT experiment. Originally, the simulation model did not account
 for imperfections in the three polarization transfers 
(\h\ to \Ctwo, \Ctwo\ to \Cone, and final \h\ to \Ctwo). 
This simulation model
 assumes perfect polarization transfers, and 
 includes experimentally-obtained \Tone\ relaxation of all spins,
applied by the program during the two waiting periods, $t_1$ and $t_2$.
The simulation yields a two-dimensional surface of expected Information
Capacity (IC) values, normalized by $\eps^2/\ln4$ as a function of both waiting
periods, as shown in Figure~\ref{fig:matlab}. Note the broad maximum region
encompassing a large extent of $t_1$ values.

It is not simple to take inevitable imperfections into account
{\em theoretically}, but it is possible to do so {\em in practice} 
by using transfer efficiency rates obtained experimentally. 
A ``practical'' simulation
model can thus account for imperfections in the three PTs. 
We obtained the results $92\% \pm 2\%, 69\% \pm 1\% $ and $74\% \pm 1\% $ 
for the three PTs, resulting in an information content of $22.4 \pm 0.9 $. 
These transfer efficiency rates
were obtained by measuring the relative peak integrals after various
intermediate stages in laboratory conditions identical to the AC experiment.

\begin{figure}[p]
{\centering \includegraphics[scale=1]{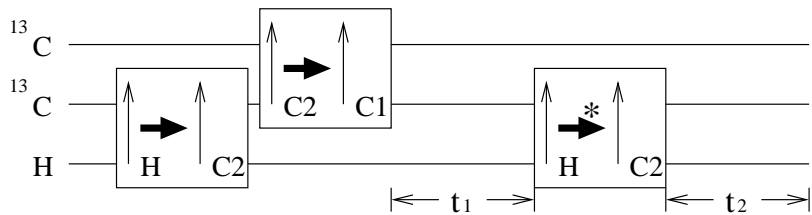} \par}
\caption{A block diagram of the complete experiment, with \Cone\
  represented at the top line.  The arrow boxes denote polarization
  transfers in the direction of the arrow. The periods $t_1$ and
  $t_2$ are the variable delay times in which we wait for \h\ to repolarize.}
\label{fig:full-experiment}
\end{figure}

% Pulse sequence and intermediate spectra
Only ``hard'', high-power, non-selective pulses were used for the
\isotope{C}{13}.  In the single case where addressing only one of them was
necessary, we induced a phase separation by employing the chemical shift,
while adequately refocusing coupling evolutions. We chose not to use
time-averaging or phase cycling, in order to study the effect of AC
independently of all other techniques.
A block diagram depicting the various stages of our experiment is
shown in Figure~\ref{fig:full-experiment}. The three transfer
sequences, from \h\ to \Ctwo, from \Ctwo\ to \Cone, and finally from \h\ to 
\Ctwo\ again, use different refocusing schemes to compensate for unwanted
couplings and chemical shifts (the \isotope{C}{13} channel was chosen to be in
resonance with \Ctwo ). In addition, the third transfer is designed to leave intact 
any polarization already stored on \Cone\ (indicated by the $*$ sign in
Figure~\ref{fig:full-experiment}). In this case, we employ refocusing to
prevent unwanted polarization transfer out of \Cone.
The first two polarization transfers jointly consist of two overlapping
refocused INEPT sequences, while the third transfer is a single refocused
INEPT. The INEPT sequences are shorter and consist of fewer pulses than the
corresponding SWAP sequences, facilitating implementation.

% Data analysis
The POTENT sequence was run with various combinations of $t_1$ and $t_2$ 
delays. To obtain statistical information several spectra were acquired 
for each combination of delays. Reported values were obtained for five 
separate single-scan measurements.

In order to validate the sequence and estimate transfer efficiencies,
truncated versions of the complete pulse sequence were acquired, each version
terminating at a different stage. Intermediate spectra obtained in this manner
are shown in Figure~\ref{fig:spectra}.

\def\figscale {0.3}
\begin{figure}[p]
{\centering
\begin{tabular}{|c|c|c|}
\cline{2-2} \cline{3-3} 
\multicolumn{1}{c|}{}&
Step&
Spectrum after the step\\
\cline{2-2} \cline{3-3} 
\hline 
1&
\resizebox*{\figscale\textwidth}{!}{\includegraphics{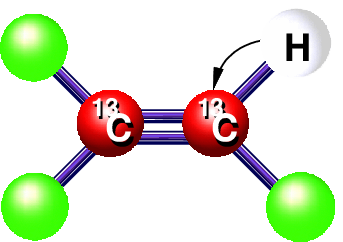}} &
\resizebox*{\figscale\textwidth}{!}{\includegraphics{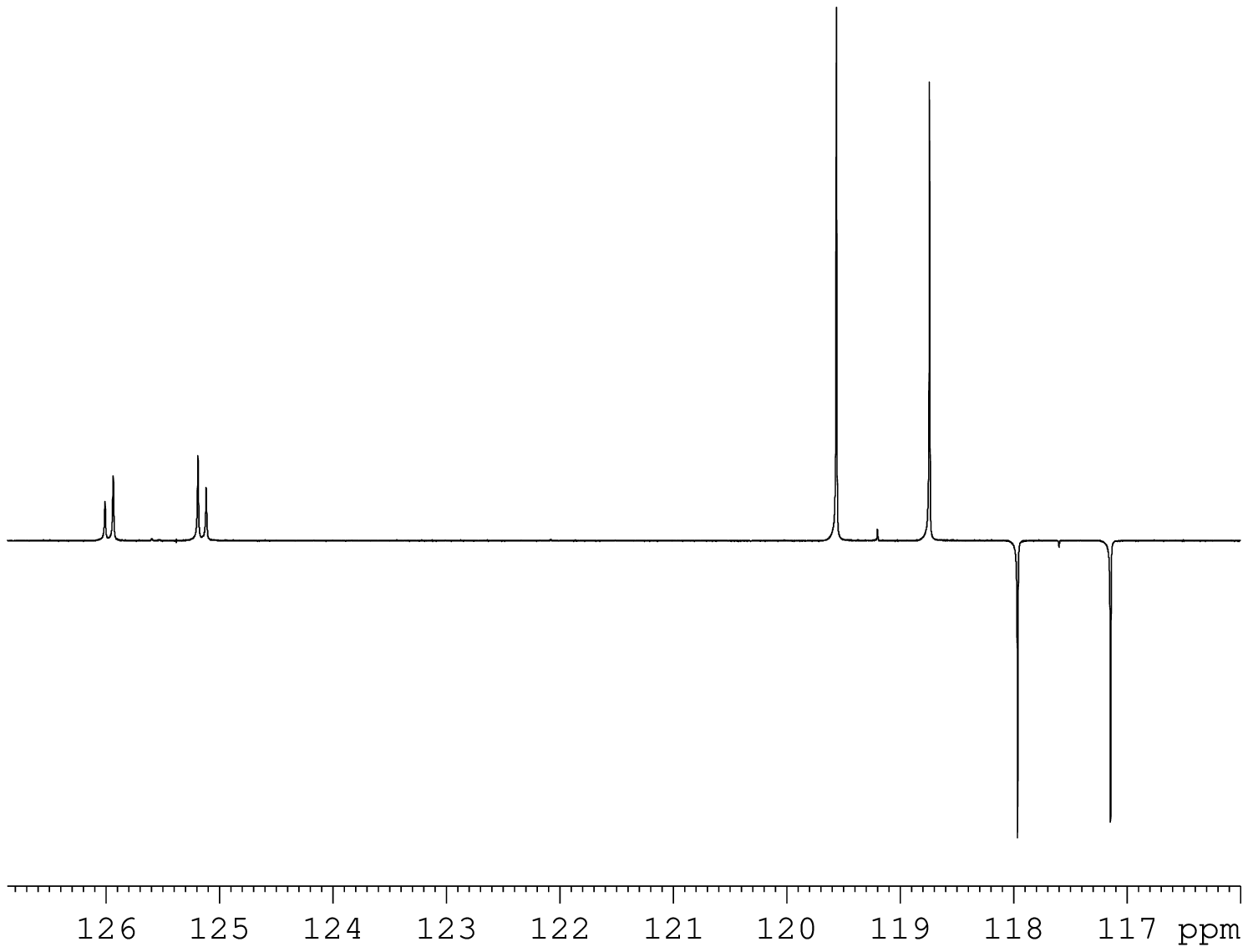}} \\
\hline 
2&
\resizebox*{\figscale\textwidth}{!}{\includegraphics{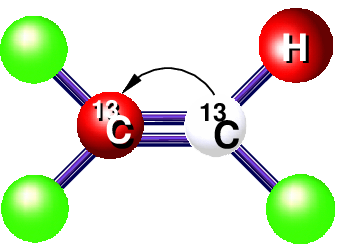}} &
\resizebox*{\figscale\textwidth}{!}{\includegraphics{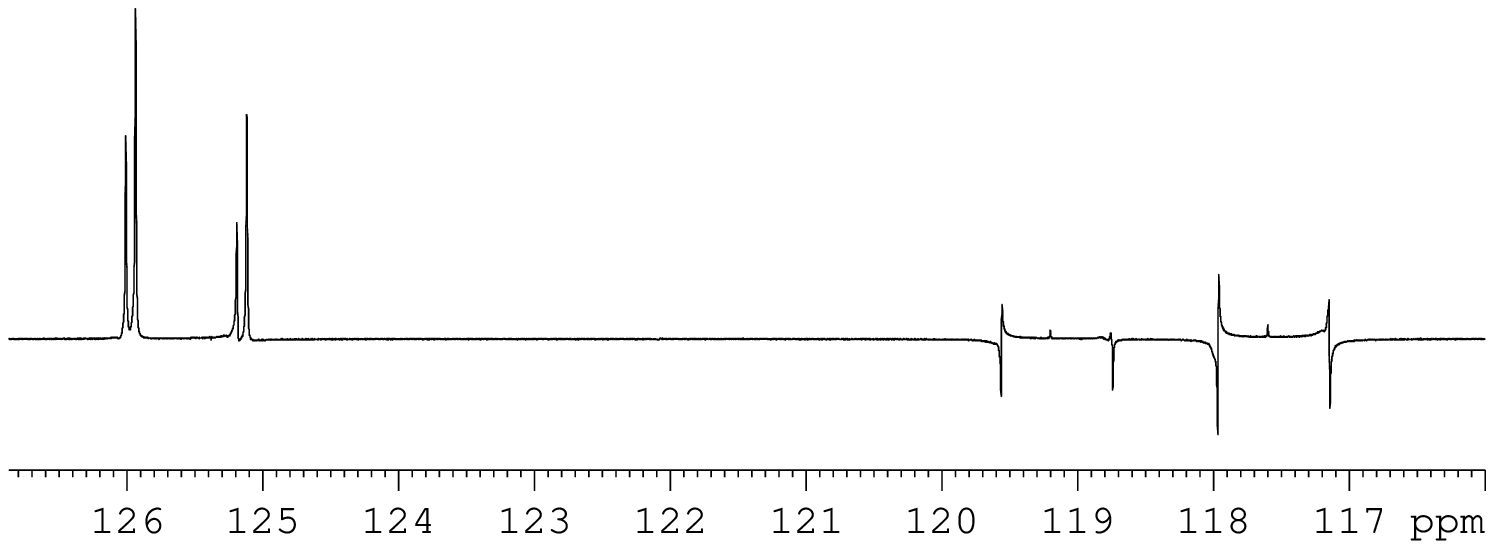}} \\
\hline
3&
\resizebox*{\figscale\textwidth}{!}{\includegraphics{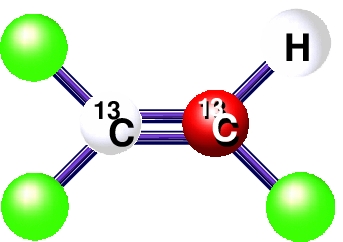}} &
\resizebox*{\figscale\textwidth}{!}{\includegraphics{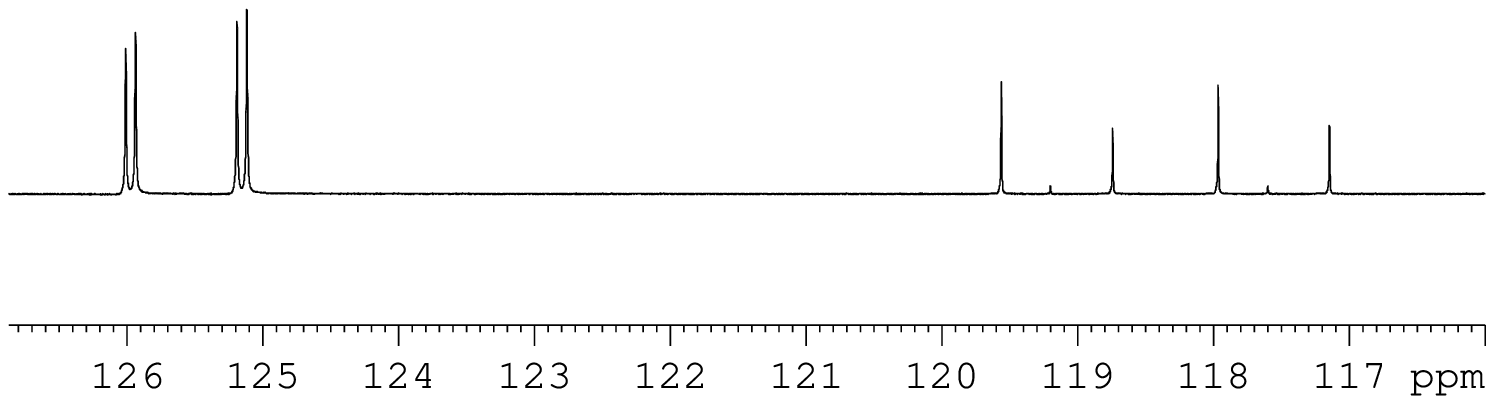}} \\
\hline 
4&
\resizebox*{\figscale\textwidth}{!}{\includegraphics{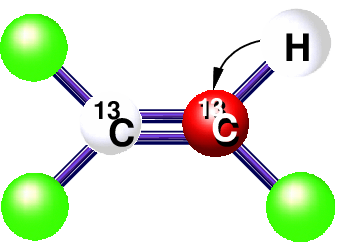}} &
\resizebox*{\figscale\textwidth}{!}{\includegraphics{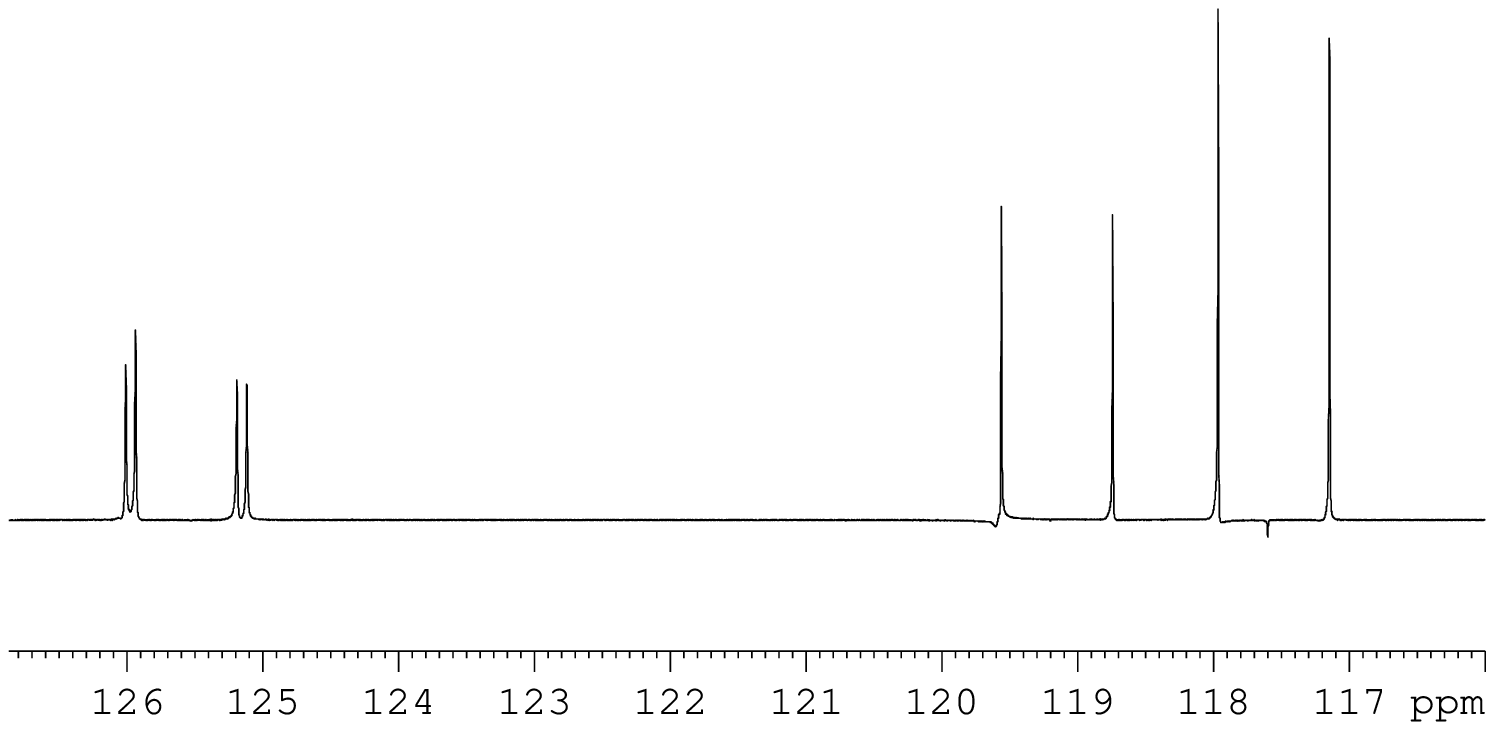}} \\
\hline
5&
\resizebox*{\figscale\textwidth}{!}{\includegraphics{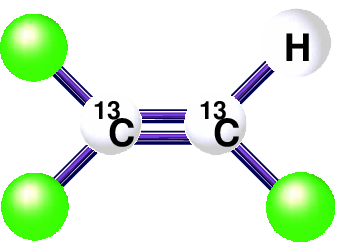}} &
\resizebox*{\figscale\textwidth}{!}{\includegraphics{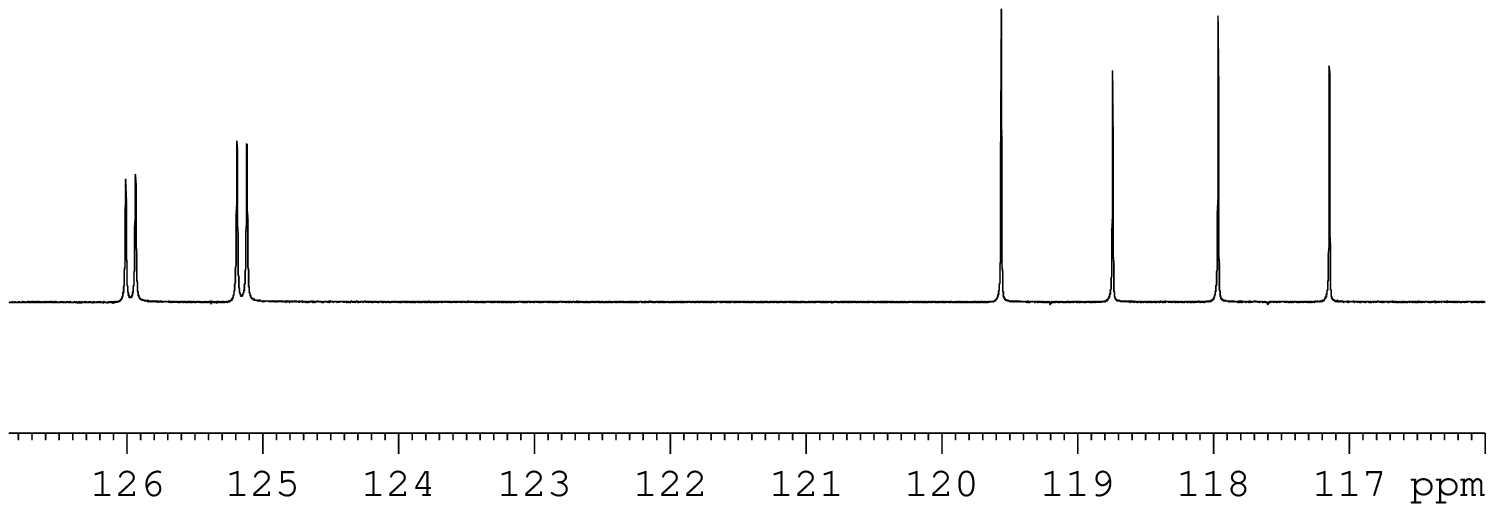}} \\
\hline
\end{tabular}\par}
\caption{The steps of the cooling experiment and the resulting
  \isotope{C}{13} spectra after each step.
}
\label{fig:spectra}
\end{figure}